\documentclass{jpp}
\usepackage{graphicx}

\usepackage{tensor}
\usepackage{bm}

\usepackage[utf8]{inputenc}
\usepackage[T1]{fontenc}
\usepackage{amsmath}

\shorttitle{Heating of Disk Coronae and Jets by GRMHD Turbulence}
\shortauthor{B. D. G. Chandran, F. Foucart, and A. Tchekhovskoy}

\title{Heating of Accretion-Disk Coronae and Jets by General Relativistic MHD Turbulence}

\author{Benjamin D. G. Chandran\aff{1} \corresp{\email{benjamin.chandran@unh.edu}}, 
Francois Foucart \aff{1,2,}
\and Alexander Tchekhovskoy \aff{2,3,4}}

\affiliation{\aff{1} Department of Physics, University of New Hampshire, Durham, New Hampshire 03824,
  USA \aff{2} Lawrence Berkeley National  Laboratory, 1 Cyclotron Rd,  Berkeley, CA 94720, USA
\aff{3} Departments of Astronomy and Physics, Theoretical Astrophysics Center, University of California Berkeley, Berkeley, CA 94720-3411, USA \aff{4} Center for Interdisciplinary Exploration \& Research in Astrophysics (CIERA),
Physics \& Astronomy, Northwestern University, Evanston, IL 60202, USA}

\begin{document}

\maketitle

\begin{abstract}
  Turbulence in an accretion disk launches Alfv\'en waves (AWs) that
  propagate away from the disk along magnetic field lines.  Because
  the Alfv\'en speed varies with distance from the disk, the AWs
  undergo partial non-WKB reflection, and counter-propagating AWs
  subsequently interact, causing AW energy to cascade to small scales
  and dissipate. To investigate this process, we introduce an
  Elsasser-like formulation of general relativistic MHD (GRMHD) and
  develop the theory of general relativistic reduced MHD in an
  inhomogeneous medium. We then derive a set of equations for the
  mean-square AW amplitude~$M_+$ and turbulent heating rate~$Q$ under
  the assumption that, in the plasma rest frame, AWs propagating away
  from the disk are much more energetic than AWs propagating toward
  the disk. For the case in which the background flow is axisymmetric
  and time-independent, we solve these equations analytically to
  determine $M_+$ and~$Q$ as functions of position. We find that, for
  an idealized thin disk threaded by a large-scale poloidal magnetic
  field, the AW energy flux is
  $\sim (\rho_{\rm b}/\rho_{\rm d})^{1/2} \beta_{\rm net,d}^{-1/2}$
  times the disk's radiative flux, where $\rho_{\rm b}$ and
  $\rho_{\rm d}$ are the mass densities at the coronal base and disk
  midplane, respectively, and $\beta_{\rm net,d}$ is the ratio
  (evaluated at the disk midplane) of plasma-plus-radiation pressure
  to the pressure of the average vertical magnetic field. This energy
  flux could have a significant impact on disk coronae and
  outflows. To lay the groundwork for future global simulations of
  turbulent disk coronae and jets, we derive a set of averaged GRMHD
  equations that account for reflection-driven AW turbulence using a
  sub-grid model.
\end{abstract}

\section{Introduction}
\label{sec:intro} 

Several types of evidence, including the observed velocities of stars
\citep{ghez05} and gas \citep{miyoshi95} near galactic centers and
gravitational-wave signals indicative of black-hole (BH) mergers
\citep{abbott16}, reveal BHs scattered throughout the visible universe
with masses ranging from several to billions of solar masses. As a BH
pulls in plasma from its surroundings, the plasma's angular momentum
causes the inflowing plasma to form a disk. If this accretion disk is geometrically thin
but optically thick (``a thin disk''), then turbulent viscosity
converts a significant fraction of the plasma's gravitational
potential energy into thermal energy that is radiated away before the
material reaches the central~BH. In part because of this, thin disks
are promising candidates for explaining much of the continuum emission
from high-luminosity active galactic nuclei (AGN) and stellar-mass BHs
in binary systems in their brighter states
\citep{shakura73,novikov73}.  Disks that are geometrically thick but
optically thin (``thick disks'') are generally much less luminous than
thin disks, because plasma in the disk can fall into the BH before it
radiates much of its thermal energy \citep{narayan94}, become
marginally unstable to convection, which suppresses mass inflow to the
central BH \citep{quataert00b}, or become gravitationally unbound and
flow outward \citep{blandford99}.  Thick disks are thought to be
present around low-luminosity BHs, such as Sagittarius~A$^\ast$ at the
center of our galaxy \citep{narayan94,quataert00b}.

BH/accretion-disk systems launch two types of outflows: non-collimated
winds, which can be mildly relativistic or non-relativistic, and
collimated, relativistic jets. The jets emanating from BH systems at
the centers of galaxies are particularly striking, because they can
span hundreds of kiloparsecs \citep{fanaroff74}. Theoretical studies
\citep[e.g.][]{blandford77} and numerical simulations
\citep{2003ApJ...599.1238D, 2004ApJ...611..977M,tchekhovskoy11} have
identified a promising mechanism for producing jets via a large-scale,
ordered magnetic field that threads an accretion disk or the event
horizon of a rotating BH.  The rotation of the disk or BH coils up the
magnetic field lines, which then act as a spring, pushing material
away from the disk along the spin axis.

Although this mechanism offers an explanation for jet formation and
acceleration, it is not yet clear how the mass outflow rates and
mechanical luminosities of jets and winds are
determined. Nor is it clear what accelerates the particles that cause
a jet, or the plasma at a jet's base, to radiate. For example, it is
unclear how to account for x-ray timing observations that indicate
that many luminous AGN contain compact coronae --- i.e.,
high-temperature, optically thin plasma --- within a few gravitational
radii of the central BH \citep{reis13}. 

Clues to these puzzles may be offered by a system much closer to
home. In one explanation for the heating and acceleration of the solar
wind, convection-driven photospheric motions shake the footpoints of
``open'' magnetic field lines (i.e., field lines that connect directly
to the interplanetary medium). This shaking launches AWs that
propagate along the magnetic field lines, through coronal holes
(open-field regions of the corona), and into the solar wind
\citep{cranmer05}. Because the Alfv\'en speed varies with distance
from the Sun, these outward-propagating AWs undergo partial non-WKB
reflection \citep{heinemann80,velli93}. Counter-propagating AWs
subsequently interact, which causes the AWs to become turbulent, which
in turn causes AW energy to cascade from large wavelengths to small
wavelengths and dissipate, heating the ambient plasma. This heating
increases the plasma pressure, which, along with the wave pressure,
accelerates the solar wind to supersonic speeds. This explanation for
the solar wind's origin is supported by numerous observational,
theoretical, and numerical studies
\citep[e.g.,][]{cranmer05,depontieu07,cranmer07,verdini07,verdini10,hollweg10,chandran11,
  perez13,vanderholst14,usmanov14,vanballegooijen16,vanballegooijen17}.
Turbulence plays a key role in this model, because the
large-wavelength AWs launched by
the Sun are damped so weakly that, without turbulence, they would
reach the distant interplanetary medium without appreciably damping or
heating the plasma \citep{barnes66}.  Wave reflection is a critical
component of the model because the Sun launches only
outward-propagating waves, and AWs interact to produce turbulence only
when there is a mix of counter-propagating AWs in the plasma rest
frame \citep{iroshnikov63,kraichnan65}.

In this paper, we explore the possibility that similar physical processes
contribute to the generation of accretion-disk coronae and jets. In particular, we consider
the fate of AWs that are launched by a turbulent accretion disk into the disk's corona and an
overlying outflow. To allow for spacetime curvature, relativistic  fluid
velocities, relativistic Alfv\'en speeds, and relativistic thermal velocities,
we work within the framework of general relativistic MHD
(GRMHD).  Previous studies have investigated the
heating of accretion-disk coronae by the reconnection of
magnetic loop structures
\citep[e.g.,][]{galeev79,uzdensky08}.  Our work
focuses on AW turbulence rather than magnetic reconnection, and open-field
regions rather than closed magnetic loops.

The remainder of this paper is organized as follows. In
Section~\ref{sec:derivation} we derive a set of equations that
describes AW propagation, reflection, and nonlinear interactions in an inhomogeneous
background flow. In Section~\ref{sec:axisymmetry} we
specialize to the case of a time-independent and axisymmetric
background and solve analytically for the mean-square AW amplitude and
turbulent heating rate as
functions of distance from the disk. In Section~\ref{sec:thin} we
apply our results to the corona and outflow overlying a thin accretion
disk in the $\alpha$-disk model \citep{shakura73,novikov73}.  In
Section~\ref{sec:subgrid} we derive a set of averaged GRMHD equations
in which AW turbulence is treated using a sub-grid model. These
equations complement the results of Section~\ref{sec:derivation} by
describing how AW turbulence influences the background flow via turbulent
heating and momentum deposition.

\section{Reflection-Driven Alfv\'en-Wave Turbulence in General Relativity}
\label{sec:derivation} 

GRMHD describes a highly conducting magnetized fluid under the assumption that the
Lorentz force vanishes for a charged particle at rest in the local
plasma frame. This assumption simplifies the source-free subset of
Maxwell's equations and the electromagnetic contribution to the
stress-energy tensor \citep[see, e.g.,][]{anile89,gammie03}. A GRMHD
fluid is described by the equation of mass conservation,
\begin{equation}
(\rho u^{\nu})_{;\nu} = 0 ,
\label{eq:cont} 
\end{equation} 
the stress-energy equation,
\begin{equation}
T\indices{^\mu^\nu_{;\nu}} = 0,
\label{eq:set} 
\end{equation} 
and the relativistic induction equation,
\begin{equation}
(b^\mu u^\nu - b^\nu u^\mu)_{;\nu} = 0,
\label{eq:ind} 
\end{equation}
where $\rho$ is the mass density, $u^{\mu}$ is the four-velocity,
\begin{equation}
T^{\mu \nu} = {\cal E} u^\mu u^\nu + \left(p +
  \frac{b^2}{2}\right)g^{\mu\nu} - b^\mu b^\nu
\label{eq:defset} 
\end{equation} 
is the GRMHD stress-energy tensor,
\begin{equation} 
b^\mu = \frac{1}{2} \epsilon^{\mu \nu \kappa \lambda}u_\nu
F_{\lambda \kappa}
\label{eq:b4} 
\end{equation} 
is the magnetic-field four vector, $b^2 = b^\mu b_\mu$,
 $F_{\lambda  \kappa}$ is the Faraday tensor divided by~$\sqrt{4\pi}$, 
$\epsilon^{\mu \nu \kappa \lambda}$ is the Levi-Civita tensor,
\begin{equation}
{\cal E} = \rho + u + p + b^2,
\label{eq:defE} 
\end{equation} 
$u$ (without indices) is the internal energy,  $p$ is the pressure, $g_{\mu\nu}$
is the metric tensor, and the units have been chosen so that the speed
of light is~1 \citep{gammie03, komissarov99}. The semicolon subscripts indicate covariant
differentiation, repeated indices are summed, and Greek indices range
from 0 to~3. The four-velocity satisfies
\begin{equation}
u^\mu u_\mu = -1,
\label{eq:u1} 
\end{equation} 
and it follows from Equation~(\ref{eq:b4}) that
\begin{equation}
u_\mu b^\mu  = 0.
\label{eq:bortho} 
\end{equation} 
The magnetic-field 3-vector is given by
\begin{equation}
B^i = b^i  u^t - b^t u^i,
\label{eq:defBi} 
\end{equation} 
where Latin indices range from 1 to~3, and
$t$ indices indicate the time component.
Equation~(\ref{eq:defBi}) can be inverted using Equations~(\ref{eq:u1})
and~(\ref{eq:bortho}) to give \citep{gammie03}
\begin{equation}
b^t = B^i u^\mu g_{i \mu} \qquad b^i = \frac{B^i + b^t u^i}{u^t}.
\label{eq:B4} 
\end{equation} 
Equation~(\ref{eq:ind}) can then be
rewritten as the two equations
\begin{equation}
\frac{1}{\sqrt{-g}} \partial_i \left(\sqrt{-g} B^i\right) = 0
\label{eq:divB} 
\end{equation} 
and
\begin{equation}
\partial_t \left(\sqrt{-g} B^i\right) = \partial_{j} \left[ \sqrt{-g}
  (B^j v^i - B^i v^j)\right],
\label{eq:ind2} 
\end{equation} 
where $\partial_\mu$ indicates differentiation with respect to
coordinate~$\mu$, $g$ is the determinant of the metric tensor, and
$v^i = u^i/u^t$ is the fluid 3-velocity \citep{gammie03}.  As
described in Section~\ref{sec:subgrid}, $b^\mu$ is the magnetic field
in the fluid frame (in the sense that is explained prior to
Equation~(\ref{eq:defb0})), while $B^i$ is the ``lab-frame'' magnetic
field when $g^{tt} = -1$.

\subsection{An Elsasser-Like Formulation of GRMHD}
\label{sec:Elsasser} 

\citet{elsasser50} reformulated non-relativistic MHD,
obtaining a set of equations that is useful for studying
AWs and AW turbulence.  We obtain an Elasser-like formulation of GRMHD
by multiplying Equation~(\ref{eq:ind}) by~$\pm {\cal E}^{1/2}$, adding the
resulting expression to Equation~(\ref{eq:set}), and then dividing
by~${\cal E}$. This yields
\begin{equation}
\left(z_{\pm}^\mu z_{\mp}^\nu  + \Pi g^{\mu
    \nu}\right)_{;\nu}
+ \left(\frac{3}{4} z_{\pm}^\mu z_{\mp}^\nu  + \frac{1}{4}
  z_{\mp}^\mu z_{\pm}^\nu + \Pi g^{\mu
    \nu}\right)\frac{\partial_\nu{\cal E}}{{\cal E}} = 0,
\label{eq:Elsasser} 
\end{equation} 
where
\begin{equation}
z_{\pm}^\mu = u^\mu \mp 
\frac{b^{\mu}}{{\cal E}^{1/2}} \qquad
\Pi = \frac{2p + b^2}{2\cal E}.
\label{eq:zpm} 
\end{equation} 

\subsection{Background Quantities, Fluctuations, and 
the Average Fluid Rest Frame}
\label{sec:AFRF}

We assume that each property of the fluid is the sum of a smoothly varying
background value plus a fluctuation,
\begin{equation}
u^\mu = \overline{ u^\mu} + \delta u^\mu \qquad b^\mu =
\overline{ b^\mu} + \delta b^\mu \qquad \mbox{ etc.}
\end{equation} 
We construct an ``average fluid rest frame'' (AFRF) at each point by
first transforming to locally Galilean coordinates and then carrying
out a Lorentz transformation that causes $\overline{ u^i}$ to vanish while leaving the 
metric in Minkowski form at that point. We define 
$\lambda$ to be the perpendicular correlation length (i.e., the
correlation length measured
perpendicular to~$\overline{ B^i}$) of the
velocity and magnetic-field fluctuations in the AFRF, and $L$ to be the characteristic length
scale of the background quantities and $g_{\mu \nu}$ in the AFRF. We assume that 
\begin{equation}
\lambda/ L \sim {\textit O}(\epsilon),
\label{eq:defepsilon} 
\end{equation}
where $\epsilon $ (without indices) is a small parameter. 
We use the notation $\langle \dots
\rangle$ to denote a volume average within a sphere of radius~$d$ in the AFRF, 
with
$\lambda \ll d \ll L$. For any vector $f^\mu$, we assume that the
following assertions lead to negligible error:
$\langle f^\mu \rangle$ is a vector,
$\langle  \langle f^{\mu}\rangle\rangle =  \langle f ^{\mu}
\rangle = \overline{ f^{\mu}}$, and
$\langle f\indices{^\mu_{;\nu}} \rangle = \langle f^\mu \rangle \indices{_{;\nu}}$,
with analogous statements for scalars and tensors. 
We note that $\overline{ u^\mu}$ is not a unit vector in the sense
of Equation~(\ref{eq:u1}) and is therefore not a four-velocity. It is
merely the local spatial average, in the AFRF, of~$u^\mu$.
The four-velocity of the AFRF is given in Equation~(\ref{eq:defu0}) below.

\subsection{General Relativistic Reduced MHD in an Inhomogeneous  Background}
\label{sec:RMHD} 

To motivate the next step in our analysis, we return for a moment to
the solar analogy. Spacecraft measurements indicate that $B^i$ and $v^i$
fluctuations in the solar wind are mostly transverse (orthogonal to
$\overline{ B^i}$) and non-compressive
\citep{klein91,horbury95,tumarsch95}. One reason for this is that the
dominant dissipation mechanism for slow magnetosonic modes and entropy
modes, turbulent mixing, causes the energy of these compressive modes
to decay on the timescale $\lambda/\delta u_{\rm rms}$, where
$\delta u_{\rm rms}$ is the rms amplitude of the velocity fluctuations
\citep{schekochihin16}. This timescale is shorter than the
energy-decay timescale for outward-propagating, non-compressive, AW
fluctuations, which is $\sim \lambda/\delta u_{\rm inward}$
\citep{iroshnikov63,kraichnan65}, where $\delta u_{\rm inward}$ is the
rms velocity fluctuation of the inward-propagating AWs. The inequality
$\lambda/\delta u_{\rm rms} \ll \lambda /\delta u_{\rm inward}$
follows from in~situ measurements \citep{bavassano00} and numerical
models \citep{cranmer05,verdini07,perez13,vanballegooijen16} that show
that outward-propagating AWs have much larger amplitudes than
inward-propagating AWs in the near-Sun solar wind. The other
compressive MHD mode, the fast magnetosonic mode, has an even smaller
amplitude in the solar wind than the slow magnetosonic mode
\citep{yao11,howes12,klein12}, in part because fast magnetosonic waves
launched by the Sun are reflected back towards the Sun by the rapid
increase in $v_{\rm A}$ between the chromosphere and corona
\citep{hollweg78}.

We conjecture that turbulence in jets and disk coronae is mostly
transverse and non-compressive (AW-like) for similar reasons. We thus consider just the
AW-like component of the turbulence by adopting the orderings of reduced MHD (RMHD),
\begin{equation}
(\delta u^2)^{1/2} \equiv (\delta u^\mu \delta u_\mu)^{1/2} \sim 
\left(\frac{\delta b^\mu \delta b_\mu}{\cal E}\right)^{1/2}\sim {\textit
  O}(\epsilon v_{\rm A})
\label{eq:ordering1} 
\end{equation} 
and
\begin{equation}
\delta \rho/\overline{ \rho} \sim \delta u/\overline{ u} \sim \delta
p/\overline{ p} \sim {\textit O}(\epsilon^2),
\label{eq:ordering1a} 
\end{equation} 
where 
\begin{equation}
v_{\rm A}^\mu \equiv \overline{ b^\mu}/\overline{\cal E}^{1/2}
\qquad v_{\rm A} = (v_{\rm A}^\mu v_{\rm A \mu})^{1/2},
\label{eq:defvA} 
\end{equation} 
and by adopting the RMHD assumption that in the AFRF $\delta z_{\pm i}
\overline{ B^i} =
0$ and $\partial_i \delta u^i = 0$. Equations~(\ref{eq:u1}) and
(\ref{eq:bortho}) and their averages
imply that, in the AFRF, ${\cal E}^{-1/2}\overline{ b_t} \sim {\cal E}^{-1/2}\delta b_t \sim
\delta u_t \sim {\textit O}(\epsilon^2 v_{\rm A}^2)$, from which it
follows that 
\begin{equation}
\overline{ z_{\pm}^\mu} \delta z_{\pm \mu} \sim \overline{
  z_{\mp}^\mu} \delta z_{\pm \mu} \sim {\textit O}(\epsilon^2
v_{\rm A}^2) \qquad \delta z_{\pm ;\nu}^\nu \sim {\textit O}(
\epsilon v_{\rm A}/L).
\label{eq:ordering2} 
\end{equation} 
As in non-relativistic RMHD \citep[see, e.g.,][]{schekochihin09}, we take the
parallel correlation length (i.e., the correlation length measured
parallel to $\overline{ B^i}$) of
$\delta z_\pm^\mu$  in the AFRF to be~$\sim {\textit O}(\lambda/\epsilon)
\sim {\textit O}(L)$,
and thus
\begin{equation}
\overline{ z_{\mp}^\nu} \partial_\nu \delta z_\pm^\mu \sim 
\delta z_{\mp}^\nu \partial_\nu \delta z_\pm^\mu \sim
{\textit O}(v_{\rm A}\delta z_\pm^\mu/L).
\label{eq:parallelorder} 
\end{equation}

Subtracting the average of Equation~(\ref{eq:Elsasser})  from
Equation~(\ref{eq:Elsasser}) and dropping terms $\ll \delta
z_\pm^\mu v_{\rm A}/L$, we obtain
\begin{equation}
\left(\delta z_{\pm}^\mu \overline{ z_{\mp}^\nu} 
+\overline{ z_{\pm}^\mu} \delta  z_{\mp}^\nu\right)_{;\nu}
+ \left(\frac{3}{4} \delta z_{\pm}^\mu \overline{ z_{\mp}^\nu}
+ \frac{3}{4} \overline{ z_{\pm}^\mu} \delta z_{\mp}^\nu
 + \frac{1}{4} \delta z_{\mp}^\mu \overline{ z_{\pm}^\nu}
 + \frac{1}{4} \overline{ z_{\mp}^\mu}\delta z_{\pm}^\nu\right)\frac{\partial_\nu{\overline{ \cal E}}}{\overline{\cal E}} = -N_\pm^\mu,
\label{eq:RMHD_Elsasser} 
\end{equation} 
where 
\begin{equation}
N_\pm^\mu = (  \delta z_\pm ^\mu \delta z_\mp^\nu + \delta \Pi g^{\mu
  \nu})_{; \nu}.
\label{eq:defN} 
\end{equation} 
The nonlinear $(\delta z_\pm ^\mu \delta z_\mp^\nu )_{; \nu}$ term
in~$N_\pm^\mu$ is nonzero only in the presence of both
$\delta z_+^\mu$ and $\delta z_-^\nu$ fluctuations, implying that
nonlinear interactions arise only between counter-propagating AW
packets, as in the non-relativistic limit
\citep{iroshnikov63,kraichnan65}. We assume that, as in
non-relativistic RMHD, the role of the $\delta \Pi$ term in
$N_\pm^\mu$ is merely to cancel out the compressive component of the
nonlinear term in the AFRF \citep{maron01}.

\subsection{Reflection-Driven GRMHD Turbulence}
\label{sec:imbalanced} 

We take the fluctuations to be statistically gyrotropic in
the AFRF, which, given Equation~(\ref{eq:ordering2}), implies that 
\begin{equation}
\langle \delta z_\pm^\mu \delta z_\pm^\nu \rangle =
\frac{1}{2} M_\pm \left( g^{\mu\nu} + \overline{  u^\mu}\:
  \overline{ u^\nu} - \frac{\overline{ b^\mu}\: \overline{
      b^\nu}}{b^2} \right)
\label{eq:defMcor} 
\end{equation} 
and 
\begin{equation}
\langle \delta z_+^\mu \delta z_-^\nu \rangle =
\frac{1}{2} R \left( g^{\mu\nu} + \overline{  u^\mu}\:
  \overline{ u^\nu} - \frac{\overline{ b^\mu} \:\overline{
      b^\nu}}{b^2} \right),
\label{eq:defRcor} 
\end{equation} 
where $M_\pm$ and~$R$ are scalars.
The quantity $\delta z_\pm^\mu$ corresponds to
AWs that propagate in the AFRF either parallel or anti-parallel to the
background magnetic field.
Because an accretion disk launches only outward-propagating fluctuations, we
assume in what follows that
outward-propagating AWs ($\delta z_+^\mu$, for concreteness) have much larger amplitudes than
inward-propagating AWs ($\delta z_-^\mu$), 
\begin{equation}
M_+ \gg M_- \qquad M_+ \gg R,
\label{eq:outward} 
\end{equation} 
as in the near-Sun solar wind and coronal holes \citep{bavassano00,cranmer05}.
We then drop terms that are $\ll \delta z_+^\mu
v_{\rm A}/L$ in Equation~(\ref{eq:RMHD_Elsasser}) to  obtain
\begin{equation}
\left(\delta z_+^\mu \overline{z_{-}^\nu}\right)_{;\nu} 
+ \left( \frac{3}{4} \delta z_+^\mu \overline{ z_-^\nu} 
+ \frac{1}{4} \overline{ z_-^\mu} \delta z_+^\nu\right)
\frac{\partial_\nu \overline{\cal E}}{\overline{\cal E}} = - N_+^\mu.
\label{eq:dzp} 
\end{equation} 
For future reference, when $M_+ \gg M_-$, to a good approximation
$\delta u^\mu = - \delta b^\mu/\overline{\cal E}^{1/2}$,
$\delta z_+^\mu = 2 \delta u^\mu$, and
\begin{equation}
\langle \delta u^2\rangle = \frac{ M_+ }{4}.
\label{eq:du2M+} 
\end{equation} 
Motivated by models of solar-wind turbulence that were reasonably
successful at explaining observations \citep{chandran09c,chandran11},
we approximate $N_\pm^\mu$ as a nonlinear damping term
\citep{dmitruk02}, setting
\begin{equation}
N_\pm^\mu = \gamma_\pm \delta z_\pm^\mu
\qquad 
\gamma_\pm = \frac{c_1 \sqrt{M_\mp}}{\lambda},
\label{eq:Npmg} 
\end{equation} 
where $c_1$ is some constant of order unity.
In taking $\gamma_\pm$ to
be $\propto \sqrt{M_{\mp}}$, we have made use of the fact that
$\delta z_\pm^\mu$ fluctuations are sheared only by $\delta z_\mp^\mu$ fluctuations.
Contracting Equation~(\ref{eq:dzp}) with $2\delta z_{+ \mu}$ and
averaging, we obtain
\begin{equation}
\overline{ z_-^\nu}  \partial_\nu M_{+} + M_+ 
\left(2 \overline{ z_-^\nu}_{;\nu}  + \frac{3 \overline{
      z_-^\nu} \partial_\nu \overline{\cal E}}{2 \overline{\cal E}}\right)
= - 2 \gamma_+ M_+.
\label{eq:zp2} 
\end{equation} 

Again dropping terms $\ll \delta z_+^\mu v_{\rm A}/L$ in
Equation~(\ref{eq:RMHD_Elsasser}), but this time choosing the lower
sign, we obtain
\begin{equation}
 \left(\overline{z_-^\mu} \delta z_+^\nu \right)_{;\nu} +\left( \frac{3}{4}\overline{ z_-^\mu}  \delta z_+^\nu +  \frac{1}{4}\delta
  z_+^\mu \overline{ z_-^\nu}\right) \frac{\partial_\nu \overline{
    \cal E}}{\overline{ \cal E}} = - \gamma_- \delta z_-^\mu .
\label{eq:zminus} 
\end{equation} 
Equation~(\ref{eq:zminus}) states that
$\delta z_-^\mu$ is
determined locally by balancing the rate at which $\delta z_-^\mu$
is produced by the reflection of $\delta z_+^\mu$
fluctuations against the rate at which $\delta z_-^\mu$
fluctuations are cascaded to small scales.
Contracting Equation~(\ref{eq:zminus})  with $\delta z_{\pm \mu}$ and
averaging, we obtain
\begin{equation}
\gamma_- R  = \frac{M_+ \overline{ z_-^\nu} \partial_\nu v_{\rm A}}{2v_{\rm A}}
\qquad
\gamma_- M_-  = \frac{R\, \overline{ z_-^\nu} \partial_\nu v_{\rm
                   A}}{2 v_{\rm A}}.
\label{eq:Mval1} 
\end{equation} 
Since $\gamma_-^2 M_- = \gamma_+^2 M_+$, Equation~(\ref{eq:Mval1}) yields
\begin{equation}
\gamma_+ = \left| \frac{\overline{ z_-^\nu} \partial_\nu
  v_{\rm A}}{2 v_{\rm A}}\right|,
\label{eq:valgammaplus} 
\end{equation} 
which does not depend on the unknown constant~$c_1$ introduced in Equation~(\ref{eq:Npmg}).

As $\delta z_+^\mu$ fluctuations propagate away from the disk, the
value of~$v_{\rm A}$ at their instantaneous location keeps changing.
Equations~(\ref{eq:zp2}) and~(\ref{eq:valgammaplus}) imply that each
time $v_{\rm A}$ changes by a factor of~$\sim 2$, a modest fraction of
the fluctuation energy cascades and dissipates. A substantial fraction
of the AW energy launched by a disk thus dissipates within several
$v_{\rm A}$ scale heights of the disk, offering a natural explanation
for the compact coronae detected in high-luminosity AGN
\citep{reis13}.  We show in Section~\ref{sec:subgrid} that the
turbulent heating rate is
\begin{equation}
Q = \frac{1}{2} \gamma_+ \overline{ \cal E} M_+.
\label{eq:Q} 
\end{equation} 
If $\overline{ u^\mu}$, $\overline{ v_{\rm A}^\mu}$, and
$\overline{ \cal E}$ are known, Equations~(\ref{eq:zp2}),
(\ref{eq:valgammaplus}), and~(\ref{eq:Q}) can be solved to
determine~$M_+$ and~$Q$.

\section{Reflection-Driven Alfv\'en-Wave Turbulence in a Stationary,
  Axisymmetric Background}
\label{sec:axisymmetry} 

We now work in the frame of the
central compact object (e.g., Boyer-Lindquist (\citeyear{boyer67})
coordinates) \nocite{boyer67}
and take
averaged quantities in this frame to be independent of time and
cylindrical angle~$\phi$.
We decompose the spatial components of
$\overline{ u^\mu}$  into poloidal and toroidal 3-vectors,
\begin{equation} 
\overline{ u^i} = u_{\rm p}^i + u_{\rm T}^i,
\label{eq:poto} 
\end{equation} 
and likewise for $\overline{ b^i}$, $\overline{ v_{\rm A}^i}$, and
$\overline{ B^i}$, where the poloidal vectors have vanishing~$\phi$
components and the toroidal vectors are proportional to the $\phi$~basis vector.
It then follows from Equation~(\ref{eq:ind2}) that \citep{mestel61}  
\begin{equation}
u_{\rm p}^i = \kappa B_{\rm p}^i,
\label{eq:mestel1} 
\end{equation} 
where $\kappa$ depends upon position.
Equations~(\ref{eq:cont}),
(\ref{eq:divB}), and (\ref{eq:mestel1}) imply that
\begin{equation}
B_{\rm p}^i \partial_i \left( \overline{ \rho} \kappa \right) = 0.
\label{eq:kapparho} 
\end{equation} 
With the use of Equations~(\ref{eq:B4}), (\ref{eq:defvA}), and (\ref{eq:mestel1}), we obtain
\begin{equation}
b_{\rm p}^i = \eta B_{\rm p}^i  \qquad v_{\rm Ap}^i = y u_{\rm p}^i,
\label{eq:etay}
\end{equation}  
where
\begin{equation} 
\eta = \frac{1 + \kappa \overline{ b^t}}{\overline{ u^t}} \qquad
y =  \frac{\eta}{\kappa \overline{\cal E}^{1/2}}.
\label{eq:defy} 
\end{equation}
In Appendix~\ref{ap:axi}, we show, given Equations~(\ref{eq:poto}) through
(\ref{eq:defy}),  that Equation~(\ref{eq:zp2}) 
can be rewritten as
\begin{equation}
(u_{\rm p}^i + v_{\rm Ap}^i)\partial_i \left( \chi
  M_+\right) = - 2 \gamma_+ \chi M_+,
\label{eq:GRHO1} 
\end{equation} 
where
\begin{equation}
\chi =
\frac{\overline{\cal E}^{3/2} (u_{\rm p} + v_{\rm Ap})^2 }{\overline{ \rho}^2 u_{\rm p}^2} .
\label{eq:defchi} 
\end{equation} 

Close to the disk, $v_{\rm A}$ increases with distance from the disk,
$\overline{ z_-^\nu} \partial_\nu \ln v_{\rm A} > 0$, and
Equations~(\ref{eq:valgammaplus}) and (\ref{eq:GRHO1}) imply that
$\chi M_+ v_{\rm A}$ is constant along a field line. Equivalently,
\begin{equation}
M_+ = M_{+\rm b} \left(\frac{\chi_{\rm b}}{\chi}\right)
\left(\frac{v_{\rm Ab}}{v_{\rm A}}\right),
\label{eq:Mp1} 
\end{equation} 
where the subscript~b indicates that the subscripted quantity is
evaluated at the base of the disk's corona on the magnetic field line
that passes through the observation point, at which the unsubscripted
$M_+$, $\chi$, and $v_{\rm A}$ terms in Equation~(\ref{eq:Mp1}) are
evaluated.  If $v_{\rm A}$ increases monotonically with increasing
distance from the disk, then Equation~(\ref{eq:Mp1}) remains valid to all
distances. On the other hand, if the value of $v_{\rm A}$ along the magnetic field line that passes through the
observation point  reaches a maximum value
$v_{\rm Am}$ at a distance $r=r_{\rm m}$ from the central~BH, then at $r>r_{\rm m}$
Equations~(\ref{eq:valgammaplus}) and (\ref{eq:GRHO1}) imply that
$\chi M_+/v_{\rm A}$ is constant along the magnetic field.  Combining
this result with Equation~(\ref{eq:Mp1}), we find that
\begin{equation}
M_+ = M_{+\rm b} \left(\frac{\chi_{\rm b}}{\chi}\right) \left( \frac{v_{\rm Ab} v_{\rm
    A}}{v_{\rm Am}^2} \right)
\label{eq:Mp2} 
\end{equation} 
at $r>r_{\rm m}$. 
If $v_{\rm A}$ progresses through an alternating series of maxima and
minima, 
then $M_+$ can be found by taking $\chi M_+ v_{\rm A}$ to be constant
along a field line between each $v_{\rm A}$ minimum and the next
maximum farther out, and taking $\chi M_+/v_{\rm A}$ to be constant
between each maximum and the next minimum. Once $M_+$ is determined,
the turbulent heating rate follows from Equations~(\ref{eq:valgammaplus}) and
(\ref{eq:Q}). Equations~(\ref{eq:Mp1}) and (\ref{eq:Mp2}) generalize previous results
on solar-wind turbulence \citep{dmitruk02,chandran09c} by allowing 
for curved spacetime, relativistic velocities, and non-zero
toroidal components and non-radial poloidal components of $\overline{ B^i}$ and $\overline{ u^i}$.

\section{Application to Coronae and Outflows above Thin Accretion Disks}
\label{sec:thin} 

As an example, we now apply our results to the corona and outflow
above a steady-state, thin accretion disk threaded by a
large-scale poloidal magnetic field. We define the coronal base 
to be the surface on which  $\beta_{\rm total} = 1$, where
\begin{equation}
\beta_{\rm total}\equiv \frac{2(p + p_{\rm rad})}{B^2},
\label{eq:betatotal} 
\end{equation} 
and $p_{\rm rad}$ is the radiation pressure. Below the coronal base
(i.e. in the disk), $\beta_{\rm total} > 1$; above the coronal base,
$\beta_{\rm total} < 1$.  The results of Sections~\ref{sec:imbalanced}
and~\ref{sec:axisymmetry} are based on the assumptions that
$M_+ \gg M_-$ and that reflection is the primary source of
inward-propagating AWs ($\delta z_-^\mu$). These assumptions are
reasonable above the $\beta_{\rm total}=1$ surface, because when
$\beta_{\rm total} < 1$ the magnetorotational instability (MRI) is
stable at wavelengths comparable to or smaller than the disk
thickness, and because the disk launches only outward-propagating
waves. These assumptions, however, are not satisfied below the
$\beta_{\rm total}=1$ surface, where the MRI generates a mix of
fluctuations propagating towards and away from the disk midplane.

\subsection{The Mean-Square AW Amplitude at the Coronal Base $M_{+\rm
    b}$}
\label{sec:Mpb} 

In the $\alpha$-disk model  \citep{shakura73,novikov73}, angular momentum transport can be viewed
as arising from a turbulent viscosity 
\begin{equation}
\nu_{\rm t} \sim v_{\rm T} l \sim \alpha c_{\rm s,d} H,
\label{eq:nut} 
\end{equation} 
where $v_{\rm T}$ is the rms amplitude of the turbulent velocity fluctuations in
the disk, $l$ is the correlation length of these velocity fluctuations,
$\alpha$ is a dimensionless constant, 
\begin{equation}
H \sim \frac{c_{\rm s,d}}{\Omega}
\label{eq:defH} 
\end{equation} 
is the disk thickness, $\Omega$ is the angular velocity of the disk, 
\begin{equation}
c_{\rm s} = \sqrt{\frac{p+ p_{\rm rad}}{\rho}}
\label{eq:cs} 
\end{equation} 
is the sound speed, and the $d$ subscript in Equations~(\ref{eq:nut}) and
(\ref{eq:defH}) indicates that $c_{\rm s}$ is evaluated at the disk midplane.
\cite{nauman15} carried out local
shearing-box simulations and found that, for Keplerian shear, the
correlation time of MRI-generated disk turbulence is
$\simeq 0.1 (2\pi/\Omega) \sim \Omega^{-1}$. This correlation time is
also comparable to the eddy turnover time in the disk, $l/v_{\rm T}$,
and thus we set
\begin{equation}
\frac{l}{v_{\rm T}} \sim \Omega^{-1}.
\label{eq:corrtime} 
\end{equation} 
Dividing the
second relation in Equation~(\ref{eq:nut}) by
Equation~(\ref{eq:corrtime}) and using Equation~(\ref{eq:defH}) to
eliminate~$H$, one obtains \citep{blackman04}
\begin{equation}
v_{\rm T}^2 \sim \alpha c_{\rm s,d}^2.
\label{eq:vt2} 
\end{equation} 
Since the mean-square velocity fluctuation is continuous
across the disk/corona boundary, Equations~(\ref{eq:du2M+}) and
(\ref{eq:vt2}) imply that
\begin{equation}
M_{+b} \sim  \alpha c_{\rm s,d}^2.
\label{eq:Mpb} 
\end{equation} 
Equation~(\ref{eq:Mpb}), in conjunction with
Equations~(\ref{eq:valgammaplus}), (\ref{eq:Q}), (\ref{eq:Mp1})
and~(\ref{eq:Mp2}), can be used to determine the approximate mean-square AW
amplitude and heating rate at all positions in the corona and outflow,
provided the dependence of $ v_{\rm A}^\nu$,
$\overline{ u^\nu}$, $\overline{ \rho}$, and $\overline{ \cal{E}}$ on
position is known.

\subsection{The AW Luminosity of a Thin Accretion Disk}
\label{sec:AWluminosity} 

To estimate the AW
energy flux from the disk, we consider the
disk's low atmosphere, in which the thermal, Alfv\'en, and poloidal outflow
velocities are non-relativistic,
$\overline{ {\cal E}} \simeq \overline{ \rho}$,
$u_{\rm p} \ll v_{\rm A p}$, and the rotational velocity is at most
trans-relativistic. For simplicity, we neglect corrections from
spacetime curvature. 
The AW contribution to the poloidal energy
flux averaged over an annulus 
of radius~$r$ and width~$\Delta r \ll r$ centered on the disk's spin axis
at height~$h$ above the coronal base is 
\begin{equation}
  F_{\rm AW} \simeq \overline{ \rho} \langle \delta u^2 \rangle v_{\rm
  Ap} f \simeq  \overline{ \rho}^{1/2} \langle \delta u^2 \rangle B_{\rm
  p, net},
\label{eq:Fpi} 
\end{equation} 
where $f$ is the fraction of the annulus that is
threaded by open magnetic field lines (that connect directly to the distant outflow/jet),
and
\begin{equation}
B_{\rm p, net} = f B_{\rm p}
\label{eq:defBpnet} 
\end{equation} 
is the azimuthally averaged poloidal magnetic flux per unit area
(i.e., the averaged vertical magnetic field) at radius~$r$ and
height~$h$.\footnote{The magnetic flux through an individual annulus
  that arises from closed magnetic loops need not vanish,
  because loops can connect one annulus to another. Nevertheless, we ignore
  the magnetic flux associated with closed loops because they
  contribute zero net flux through the disk as a whole.}  The factor
of~$f$ in Equation~(\ref{eq:Fpi}) arises because we ignore the AW
energy flux on magnetic arches or ``closed loops,'' which are rooted
at both ends in the disk. Because each magnetic loop extends only a
finite distance into the corona, $f$ is an increasing function of~$h$.
Since open magnetic field lines fan out to fill the volume above
closed loops, $B_{\rm p}$ is a decreasing function of~$h$. The product
$fB_{\rm p} = B_{\rm p, net}$, however, is approximately independent
of~$h$, because the same amount of magnetic flux passes through each
plane parallel to the disk.

In the low atmosphere, the value of $\eta$ in Equation~(\ref{eq:defy})
is $\sim 1$, and Equations~(\ref{eq:kapparho}), (\ref{eq:etay}),
(\ref{eq:defy}), and~(\ref{eq:defchi}) imply that, along a magnetic
field line, $v_{\rm Ap}/u_{\rm p} \propto \overline{ \rho}^{1/2}$, and
\begin{equation}
\chi\propto \overline{ \rho}^{1/2}. 
\label{eq:chilowcorona} 
\end{equation} 
Although AW energy dissipates in the low corona, we count such
dissipated energy as part of the AW energy flux from the disk.  To
estimate~$F_{\rm AW}$, we thus neglect dissipative losses when
determining $\langle \delta u^2 \rangle$ in Equation~(\ref{eq:Fpi}).
Equation~(\ref{eq:GRHO1}), with $\gamma_+ \rightarrow 0$, implies that
$\langle \delta u^2 \rangle \propto 1/\chi$ along a field line.
Combining this scaling with Equation~(\ref{eq:chilowcorona}), we find
that $\overline{ \rho}^{1/2} \langle \delta u^2 \rangle$ is constant
along magnetic field lines, and hence approximately independent of~$h$
in the low corona.  Since
$\overline{ \rho}^{1/2} \langle \delta u^2 \rangle$ and
$B_{\rm p,net}$ are both independent of~$h$, our estimate
of~$F_{\rm AW}$ is insensitive to the exact height at which we
evaluate Equation~(\ref{eq:Fpi}).\footnote{AWs do lose energy as the
  wave-pressure force does work on the outflowing plasma, but the
  foregoing arguments show that this energy-loss mechanism is
  negligible in the low atmosphere.}  At the coronal base,
Equation~(\ref{eq:Fpi}) can be written, with the aid of
Equations~(\ref{eq:du2M+}) and (\ref{eq:Mpb}), as
\begin{equation}
F_{\rm AW} \sim  \rho_{\rm b}^{1/2} \alpha c_{\rm s,d}^2 B_{\rm p, net},
\label{eq:FAWpb} 
\end{equation} 
where $\rho_{\rm b}$ is the density at the coronal base.
In the $\alpha$-disk model, the radiative flux from the disk is 
\begin{equation}
q \sim \alpha \rho_{\rm d} c_{\rm s,d}^3,
\label{eq:qrad}
\end{equation} 
where $\rho_{\rm d}$ is the midplane density. Upon dividing
Equation~(\ref{eq:FAWpb}) by Equation~(\ref{eq:qrad}), we obtain
\begin{equation}
\frac{F_{\rm AW}}{q} \sim  
\left(\frac{\rho_{\rm b}}{\rho_{\rm d}}\right)^{1/2} \beta_{\rm
  net,d}^{-1/2},
\label{eq:fluxratio} 
\end{equation} 
where 
\begin{equation}
\beta_{\rm net} = \frac{2(p+p_{\rm rad})}{B_{\rm p, net}^2},
\label{eq:betanet} 
\end{equation} 
and $\beta_{\rm net,d}$ is the value of $\beta_{\rm net}$ at the disk
midplane. All quantities in Equation~(\ref{eq:fluxratio}) are
functions of distance~$r$ from the central compact object.  Since~$q$
peaks at small~$r$, the ratio of the disk's AW luminosity to its
radiative luminosity is approximately equal to the right-hand side of
Equation~(\ref{eq:fluxratio}) evaluated near the disk's inner
edge.\footnote{ Here, we have assumed that the right-hand side of
  Equation~(\ref{eq:fluxratio}) increases as~$r$ decreases or
  depends more weakly on~$r$ than does~$q$.}

Although Equation~(\ref{eq:fluxratio}) in principle
determines~$F_{\rm AW}$, the factors on the right-hand side of
Equation~(\ref{eq:fluxratio}) have large uncertainties. In numerical
simulations of disks, $\rho_{\rm b}/\rho_{\rm d}$ ranges from
$\simeq 10^{-2}$ to $\simeq 0.5$, depending on a number of factors,
including whether radiation pressure dominates over plasma pressure
and whether the simulation is local or global
\citep[e.g.,][]{jiang14,jiang16,zhu18}. The value of
$\beta_{\rm net,d}$ depends on the efficiency with which a disk
accretes poloidal magnetic flux. A number of studies have argued that
if the magnetic field is passively transported by turbulence
in the disk and the turbulent diffusion coefficient is uniform, then
the time scale for poloidal flux to diffuse outward is shorter than
the accretion time scale of the matter
\citep[e.g.,][]{vanballegooijen89,livio99,guan09,guilet13}.  In this
case, little poloidal flux accumulates near the central object, and
$\beta_{\rm net, d}$ is extremely large. On the other hand, poloidal
flux may be dragged inwards much more efficiently if the turbulent
diffusion coefficient becomes small near the boundary between the disk
and its atmosphere, if the vertical 
magnetic field in the disk is concentrated
into bundles with a small volume filling factor, or if the vertical
field exerts significant torque on the disk material
\citep{livio99,spruit05,guilet13}. In numerical simulations, if the
initial magnetic field is very weak or has toroidally shaped flux
surfaces with comparatively small major radii~$r_{\rm i}$, then little
poloidal magnetic flux builds up near the central object \citep[see,
e.g.,][]{devilliers03,beckwith08,mckinney12}.  In contrast, if the
initial field is sufficiently strong and $r_{\rm i}$ sufficiently
large, or if substantial poloidal magnetic flux is continuously
injected into the simulation domain, then poloidal magnetic flux is
dragged inwards so efficiently that the outward magnetic pressure
force on disk material becomes comparable to the gravitational force,
leading to a magnetically arrested disk
\citep{igumenshchev03,tchekhovskoy11}. 

We note that the ability of turbulent heating to produce a hot corona
starting at the $\beta_{\rm total}=1$ surface depends not only
on~$F_{\rm AW}$, but also on the optical depth at this
surface,~$\tau_{\rm b}$. If $\tau_{\rm b} \gg 1$, and if radiative
transfer dominates the vertical energy flux above the
$\beta_{\rm total}=1$ surface, then the temperature must be a
decreasing function of distance from the disk midplane in order to
drive an outward radiative energy flux \citep[see, e.g., Simulation~B
of][]{jiang14}.

\subsection{Self-Consistency of the RMHD and $\alpha$-Disk Approximations}
\label{sec:selfconsistency} 

One of the assumptions underlying the reduced~MHD analysis of
Sections~\ref{sec:RMHD},
\ref{sec:imbalanced}, and~\ref{sec:axisymmetry} is the inequality
\begin{equation}
\frac{\langle \delta u^2 \rangle}{v_{\rm A}^2} \ll 1.
\label{eq:ordering3} 
\end{equation} 
Close to the coronal base, $v_{\rm A}$ increases
with distance from the disk, and we can use Equations~(\ref{eq:du2M+}) and
(\ref{eq:Mp1}) to rewrite Equation~(\ref{eq:ordering3}) as
\begin{equation}
\left(\frac{M_{+\rm b}}{v_{\rm Ab}^2}\right)\left(\frac{\chi_{\rm b}}{\chi}\right)
\left(\frac{v_{\rm Ab}}{v_{\rm A}}\right)^3 \ll 1.
\label{eq:ordering4} 
\end{equation} 
Given Equations~(\ref{eq:du2M+}) and (\ref{eq:Mpb}),
\begin{equation}
\frac{M_{+\rm b}}{v_{\rm Ab}^2} \sim \frac{\alpha c_{\rm
    s,d}^2}{B_{\rm b}^2/\rho_{\rm b}} 
\sim \alpha \beta_{\rm total,d} \left(\frac{\rho_{\rm b}}{\rho_{\rm
      d}}\right),
\label{eq:inequality} 
\end{equation} 
where $B_{\rm b}$ is the magnetic-field strength at the coronal base,
and $\beta_{\rm total, d}$ is the value of~$\beta_{\rm total}$ at the
disk midplane.
In writing the second order-of-magnitude relationship in
Equation~(\ref{eq:inequality}), we have taken
the magnetic-field strength to be fairly uniform across the
vertical profile of a disk (see, e.g., Figure~6 of
\cite{beckwith08} or Figure~2 of \cite{jiang14}).
The factor of $\alpha \beta_{\rm total,d}$ in Equation~(\ref{eq:inequality}) is typically one
over a few \citep{blackman08,sorathia12,hawley13}. Since $\rho_{\rm
  b}/\rho_{\rm d}$ is also less than~1 (or $\ll 1$),
$M_{\rm +b}/v_{\rm Ab}^2$ is smaller than~1, and
Equation~(\ref{eq:ordering4})  is at least marginally satisfied at the
coronal base. 
In the low corona, where the flows are at most
trans-relativistic and $\overline{ {\cal E}} \sim \overline{ \rho}$,
Equation~(\ref{eq:chilowcorona}) implies that
\begin{equation}
\left(\frac{\chi_{\rm b}}{\chi}\right) \left(\frac{v_{\rm Ab}}{v_{\rm
      A}}\right)^3 \sim 
\frac{\overline{ \rho}}{\rho_{\rm b}}\left(\frac{B_{\rm b}}{\overline{
      B}}\right)^3.
\label{eq:secondfactor} 
\end{equation} 
Because the density decreases rapidly with
increasing distance~$h$ from the coronal base
in the low disk atmosphere, we expect the right-hand side of
Equation~(\ref{eq:secondfactor}) to decrease with
increasing~$h$. Thus, not only is Equation~(\ref{eq:ordering4}) 
at least marginally satisfied at the coronal base, but it becomes a
better approximation in the low corona.
It is possible
that the RMHD approximation breaks down at sufficiently large distances from the
disk, but whether this happens depends upon the spatial profiles of
the background flow and
magnetic field. These profiles, in turn, depend upon how much poloidal
magnetic flux is accreted towards the central compact object, which,
as discussed above, is uncertain. 

The results of this section are also based upon the $\alpha$-disk
model, in which disk turbulence is the only mechanism for transporting
angular momentum away from the central object, and radiation is the
only means by which energy escapes from the disk surface.  If poloidal
magnetic flux is accreted efficiently towards the central compact
object and the disk drives a powerful outflow, then the angular
momentum flux and energy flux associated with this outflow could
modify the disk structure. Our use of the $\alpha$-disk model thus
becomes problematic in the most interesting parameter regime for AW
heating, in which $\beta_{\rm net,d}$ is not much greater than~1.  If
astrophysical disks reach this regime, further work will be needed to
model the disks, winds, and AW turbulence self-consistently.

\section{A Sub-Grid Model for Incorporating Reflection-Driven Alfv\'en-Wave
  Turbulence into Numerical Simulations of the Averaged GRMHD Equations}
\label{sec:subgrid} 

The correlation length of the turbulent fluctuations in a thin
accretion disk is smaller than the disk's thickness.  As a
consequence, the AWs launched into the corona of a thin disk
have a correlation length perpendicular to the magnetic field, as
measured in the AFRF in the low atmosphere of the disk, that is
smaller than the disk thickness. These AWs cascade to much smaller
scales before dissipating and heating the plasma. In order to carry
out a direct numerical simulation of the AW turbulence launched from a
thin disk into its corona and outflow, it would be necessary to
resolve, within the disk's corona and outflow, length scales much
smaller than the disk's thickness, which is unfeasible even on
today's fastest supercomputers. An alternative approach is to average
the GRMHD equations in the manner described in Section~\ref{sec:AFRF},
solve these averaged GRMHD equations numerically, and incorporate the
mean-square AW amplitude as an additional variable that evolves
according to Equations~(\ref{eq:zp2}) and (\ref{eq:valgammaplus}). In
this section, we derive a set of averaged GRMHD equations that can be
used in this way.\footnote{An analogous approach has been used in non-relativistic
simulations of the solar wind \citep{chandran11,vanderholst14,usmanov14}.}
 Throughout this section, we make use of the RMHD
orderings of Section~\ref{sec:RMHD} and neglect fluctuations in
$\rho$, $u$ (the internal energy), and~$p$.

To begin, we average
Equation~(\ref{eq:u1}) to obtain
$\overline{ u^\mu } \overline{ u_\mu}  = - (1 + \langle \delta u^2 \rangle)$.
Since $\overline{ u^\mu } \overline{ u_\mu} \neq -1$, $\overline{
  u^\mu}$ is not a four-velocity. On the
other hand, 
\begin{equation}
u_0^\mu = \frac{\overline{ u^\mu}}{\sqrt{1 + \langle \delta u^2\rangle}}
\label{eq:defu0} 
\end{equation} 
is the four-velocity of the AFRF, since its spatial components (like
those of $\overline{ u^\mu}$) vanish
in the AFRF, and since
\begin{equation}
u_0^\mu u_{0 \mu} = -1.
\label{eq:u04v} 
\end{equation} 

The mass density of the fluid, as measured by an observer with four-velocity $s^\mu$, is
$\Gamma^\mu s_\mu$, where
$\Gamma^\mu = \rho u^\mu$ is the mass-flux four vector. The averaged mass density measured by an
observer at rest in the AFRF is thus $\rho_0 = \langle \Gamma^\mu u_{0
  \mu} \rangle$,
or, equivalently,
\begin{equation}
\rho_0 = \rho \sqrt{1 + \langle \delta u^2 \rangle}.
\label{eq:defrho0} 
\end{equation} 
This mass density is larger than the rest-frame density~$\rho$
because the fluid moves with respect to the AFRF, and Lorentz
contraction causes the apparent
density of a moving fluid to be larger than the rest-frame density.

The magnetic field in the frame of an observer~O with
four-velocity~$s^\mu$ is $B_{(s)}^\mu = -(F^{\ast\mu \nu})s_\nu$,
where
$(F^{\ast\mu \nu}) = (1/2) \epsilon^{\mu \nu \kappa \tau}F_{\kappa
  \tau} = b^\mu u^\nu - b^\nu u^\mu$
is the dual of~$F^{\mu \nu}$. More precisely, $B_{(s)}^\mu$ is that
four-vector which, in the frame of observer~O, has a vanishing time
component and spatial components equal to the magnetic field that
would be measured by~O. For example, the magnetic field in the fluid
frame is $-(F^{\ast\mu \nu})u_\nu = b^\mu$. In the case that
$g^{tt} = -1$, the four-velocity $s^\mu$ of an observer moving normal
to a $t= \mbox{constant}$ ``slice'' satisfies $s_t = -1$ and
$s_i = 0$, and the magnetic field in the frame of this ``normal
observer'' is
$-(F^{\ast\mu \nu}) s_\nu = (F^{\ast\mu t}) = b^\mu u^t - u^\mu b^t
\equiv B^\mu$
\citep{duez05}.\footnote{If $g^{tt} \neq -1$, then the four-velocity
  $s^\mu$ of an observer moving normal to a $t= \mbox{constant}$ hypersurface
  satisfies $s_t = -1/\sqrt{-g^{tt}}$ and $s_i =0$. The magnetic field
  in the frame of this normal observer is then
  $ (b^\mu u^t - u^\mu b^t)/\sqrt{-g^{tt}} = B^\mu/\sqrt{-g^{tt}}$.}
The spatial components of $B^\mu$ were given in
Equation~(\ref{eq:defBi}), and $B^t$ vanishes since $F^{\ast \mu \nu}$
is antisymmetric.  The averaged magnetic field in the AFRF is
$b_0^\mu = \langle - (F^{\ast\mu \nu})u_{0\nu} \rangle$, or,
equivalently,
\begin{equation}
b_0^\mu = \overline{ b^\mu} \sqrt{1 + \langle \delta u^2\rangle} -
u_0^\mu\langle \delta u^\alpha \delta b_\alpha \rangle.
\label{eq:defb0} 
\end{equation} 
Averaging Equation~(\ref{eq:bortho}) yields 
\begin{equation}
u_0^\mu b_{0\mu} = 0.
\label{eq:bortho0} 
\end{equation} 

The continuity and induction equations for the averaged
fluid can be obtained by averaging Equations~(\ref{eq:cont}) and
(\ref{eq:ind})
and making use of
Equations~(\ref{eq:defu0}), (\ref{eq:defrho0}), and~(\ref{eq:defb0}). This yields
\begin{equation}
(\rho_0 u_0^\nu)_{;\nu} = 0
\label{eq:cont0} 
\end{equation} 
and
\begin{equation}
\left(u_0^\mu b_0^\nu - u_0^\nu b_0^\mu\right)_{;\nu} = 0.
\label{eq:ind0} 
\end{equation} 

As in Section~\ref{sec:imbalanced}, 
we restrict our analysis to the case in which AWs traveling away from
the disk have much larger amplitudes
than AWs traveling toward the disk in the AFRF. However, 
in contrast
to Section~\ref{sec:imbalanced}, we here allow either $M_+$ or $M_-$
to correspond to outward-propagating AWs, since in general $\overline{
  B^i}$ points toward the disk in some regions and away from the disk
in others. We then set
\begin{equation}
\delta u^\mu = \mp \frac{\delta b^\mu}{\overline{ \cal E}^{1/2}},
\label{eq:dudb} 
\end{equation} 
where, here and throughout the remainder of this section, the upper
sign corresponds to $M_+ \gg M_-$ and the lower sign corresponds to
$M_- \gg M_+$. Equation~(\ref{eq:du2M+}) then becomes
\begin{equation}
\langle \delta u^2 \rangle = \frac{M_\pm}{4}.
\label{eq:du2Mpm} 
\end{equation} 
Upon averaging $T^{\mu \nu}$, making use
of the RMHD orderings described in Section~\ref{sec:RMHD}, and 
dropping terms $\ll \epsilon^2 v_{\rm A}^2 \overline{\cal E}$, we obtain
\begin{equation}
\langle T^{\mu \nu} \rangle = T_0^{\mu \nu} + T_2^{\mu \nu},
\label{eq:avT} 
\end{equation}
where
\begin{equation}
T_0^{\mu \nu} = {\cal E}_0 u_0^\mu u_0^\nu + \left(p +
  \frac{b_0^2}{2}\right)g^{\mu \nu} - b_0^\mu b_0^\nu,
\label{eq:defT0} 
\end{equation} 
\begin{equation}
T_2^{\mu \nu} = {\cal E}_0 \langle \delta u^2 \rangle \left[
\left(2 - v_{\rm A0}^2 - \frac{\rho_0}{2{\cal E}_0}\right) u_0^\mu u_0^\nu 
+ \left(\frac{1 - v_{\rm A0}^2}{2}\right) g^{\mu \nu}
+ v_{\rm A0}^\mu v_{\rm A0}^\nu \pm u_0^\mu v_{\rm A0}^\nu \pm u_0^\nu v_{\rm A0}^\mu
\right],
\label{eq:defT2} 
\end{equation} 
and
\begin{equation}
{\cal E}_0 \equiv \rho_0 + u + p + b_0^2 \qquad
v_{\rm A0}^\mu = b_0^\mu/{\cal E}_0^{1/2}.
\label{eq:defE0} 
\end{equation} 
The average of Equation~(\ref{eq:set}), combined with
Equation~(\ref{eq:avT}), yields
\begin{equation}
T_{0\;\;\, ;\nu}^{\mu\nu} =  - T_{2\;\;\, ;\nu}^{\mu \nu}.
\label{eq:T0T2} 
\end{equation} 

Equations~(\ref{eq:zp2}), (\ref{eq:valgammaplus}), (\ref{eq:bortho0}),
(\ref{eq:cont0}), (\ref{eq:ind0}), and~(\ref{eq:T0T2}), along with an
equation of state $p = p(u, \rho)$, can be solved for $\rho_0$,
$u_0^\mu$, $b_0^\mu$, $u$, $p$, and~$M_\pm$ (with the proviso that if
$M_- \gg M_+$, then the plus and minus subscripts need to be
interchanged in Equations~(\ref{eq:zp2}) and (\ref{eq:valgammaplus})).
In this closed system of equations, the AW fluctuations are treated like
a fluid that co-evolves with the plasma, and
all quantities, including~$M_\pm$, vary on the length scale~$L$ of the
background flow, which greatly exceeds the perpendicular correlation
length~$\lambda$ of the fluctuations. The effects of turbulent heating
and momentum deposition are captured by the source
term~$ - T_{2\;\;\, ;\nu}^{\mu \nu}$ on the right-hand side of
Equation~(\ref{eq:T0T2}).  Since this source term contains time
derivatives, some care is needed when adding it to a GRMHD code. The
development of an appropriate numerical algorithm, however, lies
beyond the scope of this paper.

If we ignore, for the moment, dissipation of AW turbulence and
consider the ideal GRMHD equations, contracting
Equation~(\ref{eq:set})
with $u_\mu$ yields 
$( u^\nu u)_{;\nu} + p u^\nu_{;\nu} = 0$, which implies that the
specific entropy of each fluid element is an invariant 
\citep{anile89}. 
For example, if $p$ were simply $(\gamma - 1) u$ for some constant $\gamma$,
then $u^\nu \partial_\nu \ln (p/\rho^\gamma)= 0$ \citep{delzanna07}. However,
the dissipation of AW turbulence, which is modeled by the $-2\gamma_+
M_+$ term on the right-hand side of
Equation~(\ref{eq:zp2}), should lead to entropy
production, i.e., heating. To see how such heating results from 
Equation~(\ref{eq:T0T2}), we contract
Equation~(\ref{eq:T0T2}) with $\overline{ u_{\mu}}$, add
Equation~(\ref{eq:zp2}) times ${\cal E}_0/4$, and drop terms that are
$\ll \epsilon^2 {\cal E} v_A^3/L$, obtaining
\begin{equation} 
(\overline{ u^\nu} u)_{;\nu} + p \overline{ u^\nu}_{;\nu} = \frac{\gamma_\pm
\overline{ \cal E} M_\pm}{2} ,
\label{eq:heating} 
\end{equation} 
where it is $\overline{ u^\nu}$ rather than $u_0^\nu$ that appears in
Equation~(\ref{eq:heating}).
(The $\overline{ {\cal E}}$ on the right-hand side is interchangeable with
${\cal E}_0$ to the order of accuracy of the equation.)
We identify the right-hand side of Equation~(\ref{eq:heating})  as the
turbulent heating rate, which was previously stated in
Equation~(\ref{eq:Q}) for the case $M_+ \gg M_-$.

\section{Summary and Conclusion}
\label{sec:conclusion}

We investigate the propagation, reflection, and nonlinear evolution of
AWs launched from a turbulent accretion disk. We focus on open-field
regions, in which the magnetic field lines extend from the disk's
surface, through the corona, and into an overlying outflow. Working
within the framework of GRMHD, we derive a set of equations that can
be solved for the mean-square AW amplitude and turbulent heating rate
as functions of position, and we solve these equations analytically
for the case of a time-independent and axisymmetric background flow.
Applying these results to the corona and outflow above a thin
$\alpha$-disk, we show that the AW energy flux from the disk is
approximately
$(\rho_{\rm b}/\rho_{\rm d})^{1/2} \beta_{\rm net,d}^{-1/2}$ times the
disk's radiative flux, where $\rho_{\rm b}$ and $\rho_{\rm d}$ are the
densities at the coronal base and disk midplane, respectively, and
$\beta_{\rm net,d}$ is the ratio (evaluated at the disk midplane) of
plasma-plus-radiation pressure to the pressure of the average vertical
magnetic field. We also derive a set of averaged GRMHD equations that
describe the evolution of the background flow in the presence of
reflection-driven AW turbulence.

A general feature of reflection-driven AW turbulence is that, as AWs
propagate away from the disk into regions with varying values
of~$v_{\rm A}$, a significant fraction of the AW energy cascades and
dissipates each time $v_{\rm A}$ changes by a factor of~$\sim 2$.  As
a consequence, much of the AW energy launched by an accretion disk
cascades and dissipates within a few Alfv\'en-speed scale heights of
the disk. This makes AW turbulence a promising mechanism for explaining
the compact x-ray-emitting coronae that are observed around a number
of luminous~AGN.

In addition to generating compact AGN coronae, AW
heating could have a number of consequences for astrophysical disks,
coronae, and jets. AW heating increases the density scale height in a
disk's atmosphere, which enhances the mass outflow rate from the disk
and reduces the near-BH mass supply. By loading more mass onto the
magnetic field lines above an accretion disk, AW heating could
increase the mechanical luminosity of outflows driven by
large-scale magnetic forces. If AW heating deposits substantial energy
into the outflowing material far from the disk, this heating could
lead to faster outflows \citep[c.f.][]{leer80}. If AWs in the corona
(where $\beta_{\rm total} \ll 1$) mostly heat electrons
\citep{quataert98,quataert99,howes10,ressler15,ressler17}, then AW
dissipation could lead to substantial coronal emission.
Particle-in-cell simulations show that turbulence in relativistic pair
plasmas can lead to a power-law tail in the particle energy
distribution \citep{zhdankin17}, which raises the possibility that AW
turbulence in disk coronae and outflows could be an important source
of energetic particles.  Thermal conduction from the corona into a
thin disk may in some cases lead to the evaporation of the disk
\citep{meyer94} and the long-sought ``soft-to-hard'' state transition,
in which a thin disk inflates, becoming a thick disk. Conversely, if
the density of a thick disk of thickness $H \sim c_{\rm s,d}/\Omega$
increases sufficiently that radiative cooling causes the disk to
collapse parallel to the spin axis to form a thin disk, then
$\beta_{\rm net, d} \propto  \rho_{\rm d} c_{\rm
  s,d}^2/B_{\rm p, net}^2 \propto H/B_{\rm p, net}^2$
decreases during the collapse (since neither $B_{\rm p, net}$ nor
$\rho_{\rm d} H$ changes
during the collapse), and there may be a transient period (terminated,
e.g., by outward diffusion of poloidal magnetic flux) in which the AW
luminosity is strongly enhanced.

Finally, we note that in current numerical simulations of thin disks,
AW turbulence on length scales $< H$ is under-resolved and
significantly modified by numerical dissipation. Moreover, since disk
coronae are nearly collisionless, the AW
energy that is dissipated in a disk's corona is partitioned between electrons and protons in a way that
cannot be determined within the framework of MHD.  Further studies
aimed at capturing the AW heating process and its differential effects
on protons and electrons, either analytically or through the use of
sub-grid models in numerical simulations, will be needed in order to
determine the impact of AW heating on disk winds and jets.

We thank E. Blackman for helpful discussions about turbulence in
$\alpha$~disks and the anonymous referee for comments that led to improvements in
the manuscript. This work was supported in part by NASA grants NNX15AI80, NNX16AG81G,
and NNX17AI18G and  NSF grant PHY-1500041. F.~Foucart's
participation in this work was supported by NASA through Einstein Postdoctoral Fellowship grant
number PF4-150122 awarded by the Chandra X-ray Center, which is
operated by the Smithsonian Astrophysical Observatory for NASA under
contract NAS8-03060. A.~Tchekhovskoy's participation in this work was
supported by a TAC Fellowship at the University of California, Berkeley.

\appendix

\section{The Statistically Steady, Axisymmetric Case}
\label{ap:axi} 

In this appendix, we derive Equations~(\ref{eq:GRHO1}) and
(\ref{eq:defchi}) starting from Equation~(\ref{eq:zp2}) under the
assumption that the background flow is time-independent and
axisymmetric. Given this assumption, the average of
Equation~(\ref{eq:cont}) leads to 
\begin{equation}
\overline{u^\nu}_{;\nu} = - \overline{ u^\nu} \partial_\nu \ln \overline{
  \rho} = -u_{\rm p}^i \partial_i \ln \overline{ \rho},
\label{eq:divergenceu2} 
\end{equation} 
where we have neglected density fluctuations because of 
the reduced MHD orderings in Equation~(\ref{eq:ordering1a}).
With the use of Equations~(\ref{eq:divB}), (\ref{eq:mestel1}), 
(\ref{eq:kapparho}), and~(\ref{eq:etay}), we obtain
\begin{equation}
\overline{ v_{\rm A}^\nu}_{;\nu} =
\frac{1}{\sqrt{-g}}\,\partial_i\left(\frac{\sqrt{-g}\, \eta B_{\rm
      p}^i}{\overline{ \cal E}^{1/2}}\right)
= 
B_{\rm p}^i\partial_i \left(\frac{\eta}{\overline{ \cal
      E}^{1/2}}\right)
= 
y u_{\rm  p}^i \left(\partial_i \ln \eta - \frac{1}{2} \partial_i \ln
\overline{   {\cal E}}\right).
\label{eq:divergenceva2} 
\end{equation} 
Equations~(\ref{eq:etay}), (\ref{eq:divergenceu2}),
and~(\ref{eq:divergenceva2}) allow us to rewrite Equation~(\ref{eq:zp2}) 
in the form
\begin{equation}
(1+y) u_{\rm p}^i\partial_i \ln M_+ + u_{\rm p}^i\left[
- 2\partial_i \ln \overline{ \rho} + \left(\frac{3}{2} + \frac{y}{2}\right)\partial_i \ln
\overline{ \cal E} + 2y \partial_i \ln \eta\right] = - 2\gamma_+ .
\label{eq:Map1} 
\end{equation} 
To solve Equation~(\ref{eq:Map1}), we search for an integrating
factor~$\chi$ that satisfies the equation
\begin{equation}
(1+y) u_{\rm p}^i\partial_i \ln \chi = u_{\rm p}^i\left[
- 2\partial_i \ln \overline{ \rho} + \left(\frac{3}{2} + \frac{y}{2}\right)\partial_i \ln
\overline{ \cal E} + 2y \partial_i \ln \eta\right].
\label{eq:Map2} 
\end{equation} 
If we can find such an integrating factor, then we can combine
Equations~(\ref{eq:Map1}) and (\ref{eq:Map2}) and,  making use of
Equation~(\ref{eq:etay}), obtain
\begin{equation}
(u_{\rm p}^i + v_{\rm Ap}^i) \partial_i \ln(\chi M_+) = - 2\gamma_+,
\label{eq:GRHO_Ap} 
\end{equation} 
which is equivalent to Equation~(\ref{eq:GRHO1}).

To solve Equation~(\ref{eq:Map2}), we first simplify notation, defining
\begin{equation} 
A  =  u_{\rm p}^i \partial_i \ln \overline{ \rho} 
\label{eq:defA}, 
\end{equation} 
\begin{equation} 
C = u_{\rm p}^i \partial_i \ln \overline{ \cal E}
\label{eq:defC}, 
\end{equation} 
and
\begin{equation}
D = u_{\rm p}^i \partial_i \ln \eta,
\label{eq:defD} 
\end{equation} 
so that Equation~(\ref{eq:Map2}) becomes
\begin{equation}
(1+y) u_{\rm p}^i \partial_i \ln\chi =  - 2A + \left(\frac{3}{2} +
  \frac{y}{2}\right)C + 2yD.
\label{eq:Map3} 
\end{equation} 
We note that
\begin{equation}
u_{\rm p}^i \partial_i \ln y = A - \frac{C}{2} + D
\label{eq:T1} 
\end{equation} 
and 
\begin{equation}
-2A+ \left(\frac{3}{2} + \frac{y}{2}\right) C + 2yD
= \left(A - \frac{C}{2}  + D \right)(y-1) + \left(-A + C + D\right)(y+1).
\label{eq:T2} 
\end{equation} 
Substituting Equations~(\ref{eq:T1}) and (\ref{eq:T2})  into
Equation~(\ref{eq:Map3})  and dividing by $y+1$, we obtain
\begin{equation}
u_{\rm p}^i \partial_i \ln \chi = \left(\frac{y-1}{y+1}\right)u_{\rm
  p}^i \partial_i \ln y
- A + C + D,
\label{eq:Map4} 
\end{equation} 
or, equivalently,
\begin{equation}
u_{\rm p}^i \partial_i \ln\chi =  u_{\rm p}^i \partial_i
\ln\left[\frac{(y+1)^2}{y}\right]
- u_{\rm p}^i \partial_i \ln \overline{ \rho}
+ u_{\rm p}^i \partial_i \ln \overline{ \cal E}
+ u_{\rm p}^i \partial_i \ln \eta.
\label{eq:Map5} 
\end{equation} 
Equation~(\ref{eq:Map5}) can be immediately integrated to yield
\begin{equation}
\chi = \psi \times \frac{(y+1)^2 \overline{ \cal
    E}\eta}{y\overline{ \rho}},
\label{eq:chi0} 
\end{equation} 
where $\psi$ is any quantity that is constant along the lines of flow
and force:
\begin{equation}
u_{\rm p}^i \partial_i \psi = 
\frac{B_{\rm p}^i \partial_i \psi}{\kappa} = 0.
\label{eq:psiconst} 
\end{equation} 
We set  $\psi = 1/\kappa \overline{ \rho}$, which satisfies
Equation~(\ref{eq:psiconst})  given Equation~(\ref{eq:kapparho}).
Equations~(\ref{eq:etay}) and (\ref{eq:defy}) then
enable us to set $\psi\overline{ \cal E} \eta/(y \overline{ \rho})
= \overline{ \cal E}^{3/2}/ \overline{ \rho}^2$
and $(1+y)^2 = (u_{\rm p} + v_{\rm Ap})^2/u_{\rm p}^2$ in
Equation~(\ref{eq:chi0}), which yields Equation~(\ref{eq:defchi}).

\bibliography{articles} 

\bibliographystyle{jpp}

\end{document}